# High Power, Tunable, Continuous-Wave Fiber Lasers in the L-band using Cascaded Raman Amplifiers

S Arun, Vishal Choudhury, Roopa Prakash and V R Supradeepa[*]
*Centre for Nano Science and Engineering, Indian Institute of Science, Bangalore 560012, India*
[*]supradeepa@iisc.ac.in

*Abstract*— We demonstrate a high power, all-fiber, tunable laser source that can operate in the L-band region. A low power, tunable input laser is amplified with a recently proposed, high efficiency, 6[th] order cascaded Raman amplifier. The proposed system is scalable and overcomes the limitations of Erbium and Erbium-Ytterbium co-doped fiber lasers for power scaling. A tunable Erbium-Ytterbium co-doped fiber ring laser generating ~0.5W of power and tunable in the 1560-1590 nm wavelength range is utilized as the seed source. The output from the seed laser is amplified to ~24 W using 6th order cascaded Raman amplification. A high power Yb laser operating at 1117nm is used as the pump laser for driving the Raman conversions. The operating wavelength of the demonstrated laser in the eye-safe, atmospherically transparent region enables high power free-space applications. In addition, this source enables other interesting applications such as high power supercontinuum generation with conventional silica fibers.

## I. INTRODUCTION

Fiber lasers with their power scaling capability and superior beam quality, have become a preferred choice among high power laser technologies. They have outperformed several other existing high power laser technologies and are extensively used for material processing and directed energy applications. In addition, fiber lasers are being explored for a wide range of applications like remote sensing, free space optical communication and supercontinuum generation.

Continuous wave (CW) lasers operating in the eye-safe and atmospheric transparency wavelength region [1] are of great interest in material processing and in applications like free space optical communications [2] and remote sensing as they offer very low free space propagation loss. These applications demand highly bright, directed beam of light sources at high powers. In addition to this, they are required in other applications such as high power supercontinuum generation using conventional silica fibers [3]. The existing CW fiber lasers that work near this wavelength region are based on Erbium doped fibers (Erbium only or Erbium-Ytterbium co-doped). However, power scaling with single mode operation in Erbium lasers or Erbium co-doped with Ytterbium (EYDF) lasers is found to be unfavorable mainly because of the high quantum defect when pumped at the conventional wavelength of 976nm.

This results in thermal issues due to high quantum defect and beam quality degradation. Er-Yb fiber lasers have another serious limitation of parasitic lasing by Ytterbium ions at higher powers [4]. A laser source that can operate in this wavelength region which can deliver high average power, is tunable and ensure single mode operation is strongly desired. Cascaded Raman fiber lasers provide an elegant and convenient way to generate these high power lasers at these wavelengths.

In this work, we demonstrate a high power, fiber laser architecture which can operate in the L-band region (1565nm to 1610nm) based on the principle of cascaded Raman amplification [5], and is widely tunable over this band. We use an Ytterbium laser at 1117 nm as the pump for the recently proposed, high efficiency cascaded Raman conversion [6]. The tunable seed signal is generated using a Er-Yb codoped fiber based ring laser. We demonstrate the architecture through a ~24W laser continuously tunable from 1565nm to 1590nm. The demonstrated tuning range is primarily limited by the tuning range of the seed source. By overcoming this, extended tuning across the L-band can be obtained with the same laser. To the best of our knowledge, this is the first time, a high power amplifier at the 20W level is demonstrated in the L-band based on a high order, cascaded Raman amplifier.

## II. SYSTEM

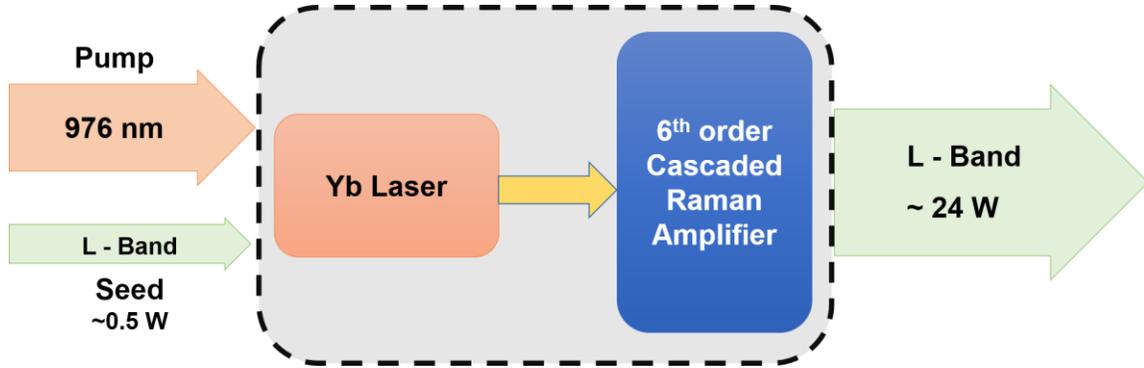

Fig 1: Schematic of the laser source

The laser architecture can be broadly divided into two parts, a seed laser and an amplifier (schematic shown in Fig 1). As a combined entity, the input-output parameters are the same as that of the primary competing technology - Erbium and Erbium-Ytterbium fiber amplifiers. A tunable seed in the L-band together with multimode pump diodes at 976nm is the input. The multimode pump diodes pump a Yb doped fiber laser at 1117nm which then performs $6^{th}$ order cascaded Raman amplification [6] of the input seed source. The tunability of the seed source enables tunability of the output.

It is important here to point to two alternate techniques developed recently, to achieve tunable, power scalable, high power sources in this wavelength region. The first technique utilizes generation of a fixed wavelength, high power source at the in-band pump wavelength of Erbium fibers (1480nm) based on Raman lasers and then core pumping Erbium fibers together with a seed [7]. This overcomes most of the power scaling problems faced when cladding pumped at 976nm. This however, introduces additional complexity in the form of an additional rare-earth doped amplifier stage and compared to this approach, the proposed work achieves similar results in a single, all passive amplifier stage. The other technique utilizes a tunable Ytterbium doped fiber laser and using the recent idea of Random Raman fiber lasers, tunability in a variety of wavelength bands is achieved [8, 9]. However, in this approach, two issues exist – Firstly, the temporal, spectral properties of the output are not user controllable. Thus this approach can only be used for CW lasers. The proposed system is an externally seeded amplifier and there is greater user control of the temporal and spectral features of the output. Secondly, tuning of the output power is limited in the previous approach. This is due to output power being intimately connected to the power in the Yb doped input laser and its variation results in affecting the cascaded Raman conversion which degrades the output power and spectral purity of the output wavelength.

## III. MECHANISM

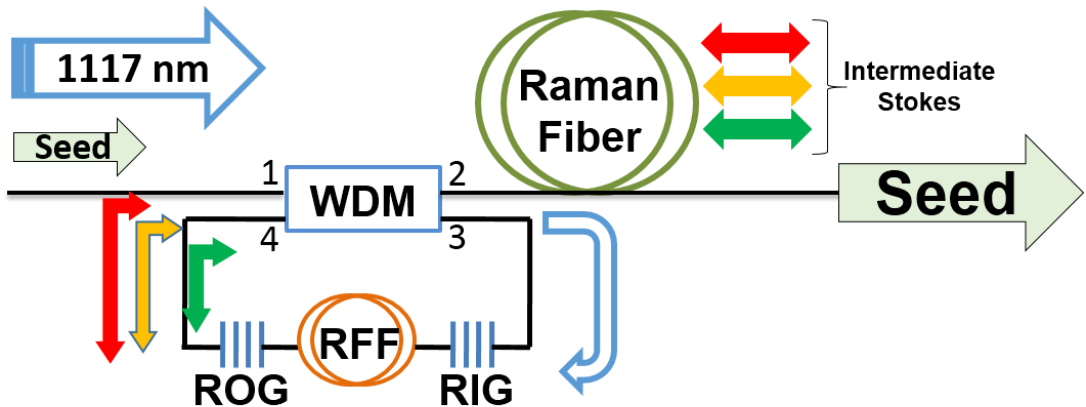

Fig 2: Detailed schematic, ROG, RIG – Raman output grating set, Raman input grating set, WDM – wavelength division multiplexer, RFF – Raman filter fiber

Conventional cascaded Raman lasers used a nested resonator configuration comprising high Raman nonlinearity fiber (referred to Raman fiber) enclosed between series of high reflectivity fiber Bragg grating pairs at all the intermediate Raman Stokes wavelengths and an output coupler at the desired output wavelength. This achieved wavelength conversion of a high power Yb doped fiber laser. This approach however suffered from lower efficiency due to losses inherent to the resonator configuration. In addition, the cascaded resonator configuration prevents external input signal for cascaded Raman amplification. Recently, a high efficiency Raman laser architecture was proposed [10, 11] in which highly efficient wavelength conversion has been demonstrated from 1117 nm to 1480 nm with ~64% efficiency (quantum limited efficiency of ~75%). This was enabled by a single pass cascaded Raman amplifier seeded at all the intermediate Raman Stokes wavelengths and a specialty fiber referred to as Raman filter fiber (RFF). In this configuration, excess losses arising from the resonator are avoided. The Raman filter fiber, based on a W-shaped index profile fiber behaves like a Raman fiber with a strong short-pass filtering function built into it. The wavelength cut-off of this filter fiber is chosen just beyond the output (final) wavelength. This prevents further Raman scattering of the output to the next undesirable order. However, in this architecture, another cascaded Raman laser (of low power) is required to generate the intermediate Raman Stokes wavelengths, which demanded substantial additional components and increased the complexity. This was overcome recently in [6], where a small fraction of the input power is tapped into a secondary Raman cavity to generate the required intermediate Raman Stokes wavelengths and then recoupled back to seed the cascaded Raman conversion. This cascaded Raman convertor forms the primary component of the proposed laser.

Figure 2 shows the detailed schematic of the proposed laser. High power light from a Yb doped fiber laser (1117nm in this experiment) is coupled with a tunable, low power L-band seed laser and then sent through Raman fiber for cascaded conversion. In addition, by tapping out a small fraction of the Yb light (through the leakage characteristics of a practical fused fiber WDM component), intermediate Stokes components are generated using a secondary cascaded Raman resonator and then recombined through the WDM. The cascaded Raman resonator and the Raman filter fiber in the secondary cavity ensure that seeding is only generated until the pen-ultimate Raman Stokes order (5$^{th}$ order, 1480nm for 1117nm input). The tunable seed which was coupled into the fiber is then Raman amplified to high powers by the pen-ultimate Stokes wavelength (1480nm).

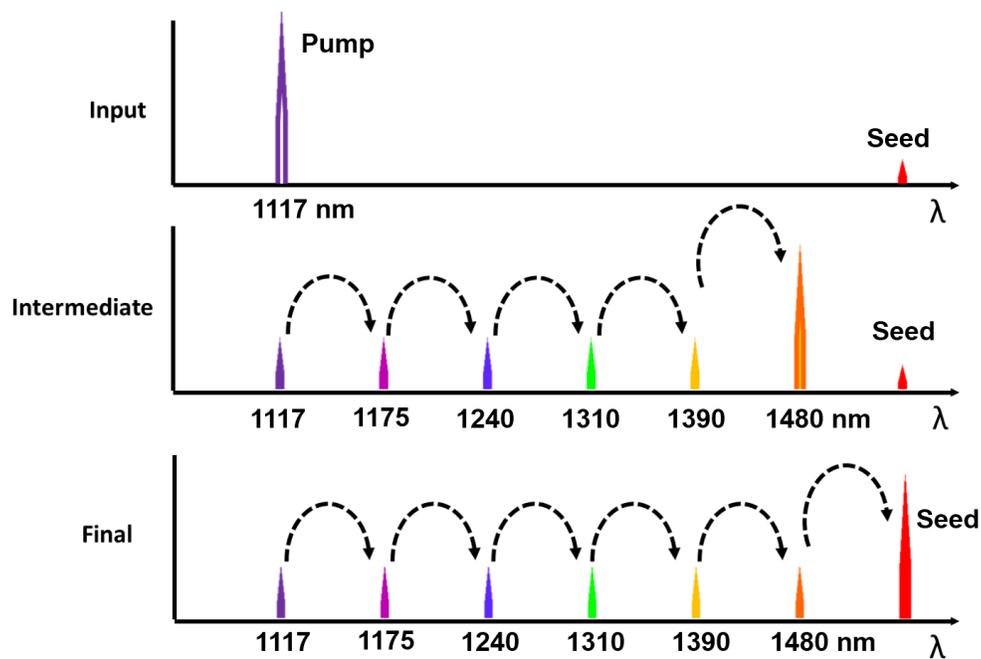

Fig 3: Schematic on power transfer among successive Stokes lines

The broadband nature of Raman gain in Silica fibers enables amplification of a widely tunable input seed. Fig 3 shows the schematic of the operation of the laser at full power through the variation of wavelengths propagating in the Raman fiber along its length. In this scheme, a fraction of power is anticipated to be left over in the intermediate Stokes wavelengths. Most of this unconverted power will be at the penultimate wavelength (unconverted pump, 1480nm in this case). This important addition also distinguishes this laser with regard to power tuning from tunable lasers based on Random Raman feedback of a tunable Yb doped laser [8, 9]. In this approach, the Yb doped fiber laser pumping the conversion is always operating at full power and with an optimized cascaded Raman amplifier for this power level, the wavelength conversion to the penultimate Stokes is efficient with very small amounts of power in all the intermediate Stokes components. Since the pump filter extracts out the penultimate stokes, in the absence of a seed, there is negligible light in the L-band (signal). By modification of the power of the seed source, output power tuning of the L band signal can be achieved.

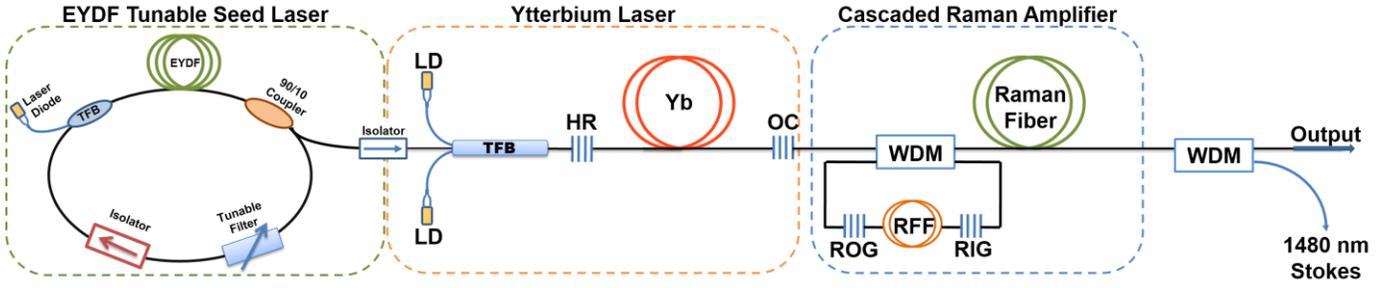

Fig 4: Detailed architecture of the high power laser source, HR/OC – High/Low reflecting grating, RIG/ROG – Raman Input/Output grating set, WDM – Wavelength division multiplexer, LD – Laser Diode, TFB – Pump signal combiner based on tapered fiber bundle

Figure 4 shows the detailed schematic of our laser. In this work, we use a Yb doped fiber laser generating ~90W at 1117nm as the high power laser source. The Yb laser architecture consists of ~20 m of single-mode, double clad Yb doped fiber (Nufern SM-YDF-5-130) spliced within a cavity formed by a pair of fiber Bragg gratings and pumped at 976 nm using laser diodes. The conversion efficiency from 976 nm to 1117 nm is ~70%. We perform cascaded Raman conversions in ~360 m of high dispersion ( > -80ps/nm/km) low effective area (12 sq micron mode field area) fiber (called as Raman fiber). The Raman fiber has a V-number of ~1.4 in the L-band (core diameter 4micron, NA ~0.2), making it a highly single-moded fiber. The 1117 nm output from Yb laser is fed into the Raman fiber through a 1117/1480 nm fused fiber WDM. Most of the input power is coupled to the straight port of WDM but, because of the inherent crosstalk associated with a practical WDM, a small fraction (~ 3%) of the input power is coupled into the cross port. We utilize this cross coupled power to generate the seed signals which is required for the intermediate Raman stokes lines in the primary, cascaded Raman conversion stage (within the Raman fiber). Seeding the intermediate stokes lines with sufficient power is essential in CW Raman processes as it improves the conversion efficiency and also helps in preferential forward scattering [10]. The low power (~3W) that is tapped out from the forward port into the cross port is then fed into a cascaded Raman resonator which is formed between using Raman filter fiber (with cut-off at 1500 nm) and Raman input and output grating sets. The Raman resonator is designed to generate seed signals at all intermediate Raman stokes lines till 1480 nm. A fraction of the power at all stokes wavelengths is then cross coupled into the straight port of the WDM where it seeds all the intermediate Raman conversions that happen in the primary, cascaded Raman amplifier.

Along with the output from the Yb laser, we also feed the seed signal from the Er-Yb laser (which is to be amplified) into the Raman amplifier. The seed laser consists of an Er-Yb co-doped fiber as the gain medium in a ring laser configuration, capable of generating an output power from 0 to ~500 mW and tunable over a bandwidth of ~30nm from 1560nm – 1590nm with a linewidth of 1nm (schematic shown in Fig 4). A crucial requirement is the feeding of the seed signal into the amplifier stage together with the output from the Yb laser. Using extra components like a coupler to feed the seed into the amplifier will introduce additional losses and complexity. This problem was solved without adding any extra optics/components by feeding the seed signal through the signal port of the pump combiner (TFB) in the Yb laser. Since the seed wavelength is outside the gain-bandwidth of the Yb gain medium, it doesn't interact with the Yb laser while it propagates in the core. By doing so the seed signal also co-propagates along with the 1117 nm pump into the Raman fiber and when the fifth order cascaded Raman conversion of the pump happens within the fiber, high power in 1480 nm wavelength is generated and with further stimulated Raman scattering along the length of the fiber, the seed signal gets preferentially amplified to high powers using the 1480 nm pump.

## IV. RESULTS

Figure 5 shows the results of our system. Fig 5(a) shows the maximal output power as a function of wavelength. A continuous tuning range from 1565nm to 1590nm was achieved. The tuning in the longer wavelength side was limited by the span of the seed laser source which was limited to 1590nm. Over 20W of output power was achieved from 1570nm to 1590nm with the maximum of ~24W. At 1565nm, ~18W of power was achieved. This reduction in output power was due to reduced Raman gain for a 1480nm pump at this wavelength. The amplifier was found to have a constant gain of ~17 dB across its wavelength range with ~27% efficiency (w.r.t power at 1117nm and ~20% overall (i.e w.r.t 976nm pumping)). This system performs comparably with the other competing technology of cladding pumped Erbium-Ytterbium co-doped fiber lasers without the associated problems of parasitic lasing and reduced mode quality [4, 15, 16]. With additional improvements on splice losses in the system and optimizing the leakage losses of the WDMs, we believe further improvement of efficiency can be obtained. Fig 5(b) shows the output spectra of the laser normalized to output power. The FWHM of the output lines was ~3nm across the tuning range. We also notice some structure in the line-shapes due to nonlinear spectral broadening. Reducing the linewidth of the seed source can contribute to reducing output linewidth.

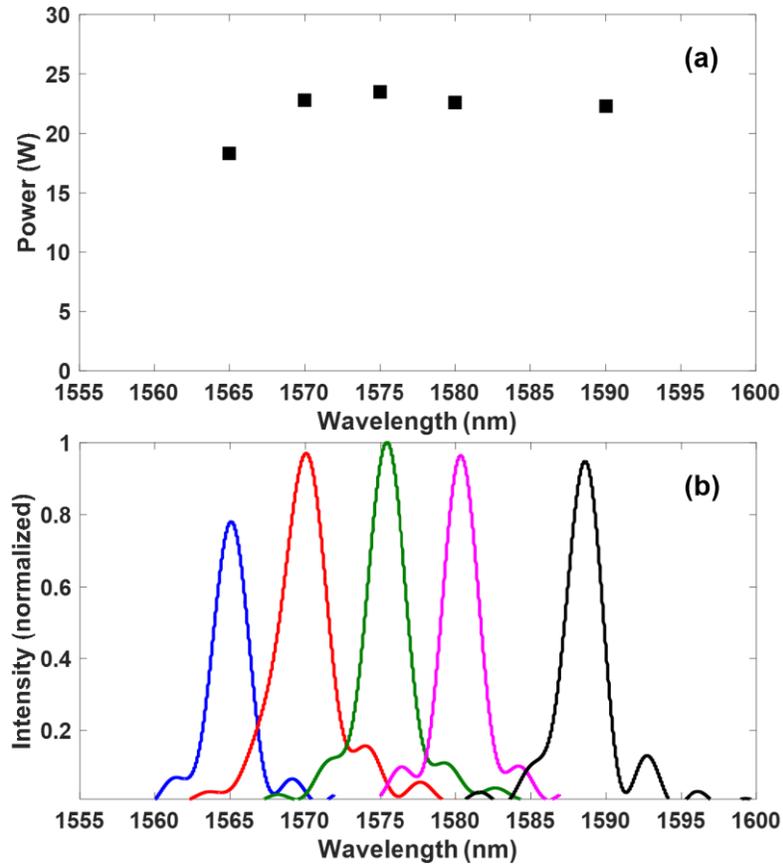

Fig 5: (a) Output power (b) Output spectrum (normalized w.r.t output power)

From the shape of Raman gain spectrum in silica-optical fibers, our models (based on equal Raman gain points on either side of the maximum at 13.2THz) indicate that over 15W of power can be obtained from the same laser until 1610nm. We further anticipate that the enhancement of the length of Raman fiber will also extend the tuning range of the laser further. This will however be at the cost of lower output power in the central region. The limitation of seed wavelength in our setup was due to the tunable filter in the ring laser. By replacing this with a longer tuning range filter, we anticipate that this laser system can be made continuously tunable in the L-band from 1565nm to 1610nm at output power levels of >15W.

## V. Summary

In this work we demonstrated a 6$^{th}$ order cascaded Raman amplifier based system to obtain high power tunable CW lasers in the L-band. With the same architecture, other input signals such as pulses, data modulated bit streams etc can be utilized. Due to the atmospheric transparency in the L-band, this would be very interesting for high bandwidth, free space optical communications.

Another interesting application for such a laser is to use it as a pump laser for supercontinuum generation in highly non-linear fibers (HNLF) [12]. For fiber based supercontinuum generation it is very important to have a high power pump laser source operating near the zero dispersion wavelength (ZDWL) of the fiber [13]. And for most conventional HNLFs, the ZDWL is in the C or L band region where high power laser sources are very limited. However, with the laser source that we have proposed here, it is possible not only to pump at the right ZDWL, but also to optimize the supercontinuum by tuning around the ZDWL of the fiber [14].